\documentclass[twocolumn,unsortedaddress,superscriptaddress]{revtex4}
\usepackage{graphicx}
\usepackage{color}
\usepackage{amsmath}
\usepackage{ulem} 
\usepackage{bm} 
\usepackage{comment}
\usepackage[breaklinks=true,colorlinks,citecolor=blue,linkcolor=blue,urlcolor=blue]{hyperref}

\begin{document}

\title{The ultrafast dynamics and conductivity of photoexcited graphene at different Fermi energies}

\author{A.\ Tomadin}\thanks{Equal contribution} \affiliation{Istituto Italiano di Tecnologia, Graphene Labs, Via Morego 30, I-16163 Genova, Italy}
\author{S.\ M.\ Hornett}\thanks{Equal contribution} \affiliation{School of Physics, University of Exeter, Stocker Road, Exeter EX4 4QL, UK.}
\author{H.\ I.\ Wang} \affiliation{Institute of Physics, Johannes Gutenberg University Mainz, 55099 Mainz, Germany} \affiliation{Max Planck Institute for Polymer Research, Ackermannweg 10, Mainz 55128, Germany}
\author{E.\ M.\ Alexeev} \affiliation{University of Sheffield, Sheffield S3 7RH, UK}
\author{A.\ Candini} \affiliation{Centro S3, Istituto Nanoscienze - CNR, via Campi 213$/$a 41125 Modena, Italy}
\author{C.\ Coletti} \affiliation{Center for Nanotechnology Innovation @ NEST, Istituto Italiano di Tecnologia, Piazza San Silvestro 12, 56127 Pisa, Italy} \affiliation{Istituto Italiano di Tecnologia, Graphene Labs, Via Morego 30, I-16163 Genova, Italy}
\author{D.\ Turchinovich} \affiliation{Max Planck Institute for Polymer Research, Ackermannweg 10, Mainz 55128, Germany} \affiliation{Fakult\"{a}t f\"{u}r Physik, Universit\"{a}t Duisburg-Essen, Lotharstr. 1, 47057 Duisburg, Germany}
\author{M.\ Kl\"{a}ui} \affiliation{Institute of Physics, Johannes Gutenberg University Mainz, 55099 Mainz, Germany}
\author{M.\ Bonn} \affiliation{Max Planck Institute for Polymer Research, Ackermannweg 10, Mainz 55128, Germany}
\author{F.\ H.\ L.\ Koppens} \affiliation{ICFO - Institut de Ci\`{e}ncies Fot\`{o}niques, The Barcelona Institute of Science and Technology, Castelldefels (Barcelona) 08860, Spain} \affiliation{ICREA - Instituci\'o Catalana de Re\c{c}erca i Estudis Avancats, 08010 Barcelona, Spain}
\author{E.\ Hendry} \affiliation{School of Physics, University of Exeter, Stocker Road, Exeter EX4 4QL, UK.}
\author{M.\ Polini} \affiliation{Istituto Italiano di Tecnologia, Graphene Labs, Via Morego 30, I-16163 Genova, Italy}
\author{K.\ J.\ Tielrooij} \email{Correspondence: klaas-jan.tielrooij@icfo.eu, andrea.tomadin@mac.com} \affiliation{ICFO - Institut de Ci\`{e}ncies Fot\`{o}niques, The Barcelona Institute of Science and Technology, Castelldefels (Barcelona) 08860, Spain}

\begin{abstract}
For many of the envisioned optoelectronic applications of graphene it is crucial to understand the sub-picosecond carrier dynamics immediately following photoexcitation, as well as the effect on the electrical conductivity -- the photoconductivity. Whereas these topics have been studied using various ultrafast experiments and theoretical approaches, controversial and incomplete explanations have been put forward concerning the sign of the photoconductivity, the occurrence and significance of the creation of additional electron-hole pairs, and, in particular, how the relevant processes depend on Fermi energy. Here, we present a unified and intuitive physical picture of the ultrafast carrier dynamics and the photoconductivity, combining optical pump -- terahertz probe measurements on a gate-tunable graphene device, with numerical calculations using the Boltzmann equation. We distinguish two types of ultrafast photo-induced carrier heating processes: At low (equilibrium) Fermi energy ($E_{\rm F} \lesssim$ 0.1 eV for our experiments) broadening of the carrier distribution involves interband transitions -- interband heating. At higher Fermi energy ($E_{\rm F} \gtrsim$ 0.15 eV) broadening of the carrier distribution involves intraband transitions -- intraband heating. Under certain conditions, additional electron-hole pairs can be created (carrier multiplication) for low $E_{\rm F}$, and hot carriers (hot-carrier multiplication) for higher $E_{\rm F}$. The resultant photoconductivity is positive (negative) for low (high) $E_{\rm F}$, which originates from the effect of the heated carrier distributions on the screening of impurities, consistent with the DC conductivity being mostly due to impurity scattering. The importance of these insights is highlighted by a discussion of the implications for graphene photodetector applications.
\end{abstract}

\maketitle

\section*{Introduction}

The potential for technological breakthroughs employing graphene is particularly high in the field of optoelectronics, given graphene's promising properties for THz technologies \cite{Crassee2011, Lee2012, Sensale2012, Vicarelli2012, Tamagnone2014, Mics2015, Tamagnone2016}, photodetection \cite{Koppens2014}, plasmonics \cite{Polini2008, Jablan2009, Grigorenko2012}, light harvesting \cite{Bonaccorso2015}, data communication \cite{Xia2009, Mueller2010, Gan2013, Pospischil2013, Schall2014}, ultrafast laser technologies \cite{Bao2009, Sun2010, Chakraborty2016}, and more. For all such applications it is of primary importance to accurately understand how photoexcitation and the subsequent ultrafast carrier dynamics occur and affect graphene's conductivity. Here, the \textit{ultrafast carrier dynamics} correspond to to the \textit{initial energy relaxation} of photoexcited electron-hole pairs, which occurs on a sub-picosecond timescale, whereas the graphene \textit{DC photoconductivity} is related to how \textit{momentum scattering} takes place in the quasi-equilibrium state after initial energy relaxation.
\

The ultrafast energy relaxation dynamics of graphene have been studied experimentally using optical pump -- optical probe spectroscopy \cite{Lui2010, Breusing2011, Brida2013, Plotzing2014, Mittendorff2014, KonigOtto2016}, time-resolved ARPES measurements \cite{Johannsen2013, Gierz2013}, time-resolved photocurrent scanning microscopy \cite{Graham2013, Tielrooij2015, Tielrooij2017}, optical pump -- terahertz (THz) probe spectroscopy \cite{George2008, Strait2011, Boubanga2012, Docherty2012, Tielrooij2013, Frenzel2013, Jnawali2013, Shi2014, Frenzel2014, Jensen2014, Hafez2015, Mihnev2016, Wang2017} and high-field THz spectroscopy \cite{Mics2015, Razavipour2015}. These time-resolved studies have identified that the energy relaxation dynamics of the photoexcited carriers consist mainly of carrier-carrier scattering and coupling to optical, acoustic and remote substrate phonons. Carrier-carrier scattering leads to a very rapid ($t \lesssim $ 50 fs) broadening of the electron distribution as a function of energy, associated with thermalization within the electronic system, i.e.\ carrier heating \cite{George2008, Lui2010, Breusing2011, Brida2013, Tielrooij2013, Johannsen2013, Gierz2013, Gierz2015}. Some theoretical studies \cite{Winzer2010, Girdhar2011, Winzer2012, Kadi2015, Malic2016}, as well as combined experimental-theoretical work \cite{Brida2013, Gierz2013, Plotzing2014, Gierz2015} have discussed the interband scattering processes that lead to carrier multiplication (CM), where the carrier density in the conduction band exceeds the carrier density immediately after photoexcitation. Other studies have discussed the occurrence of intraband scattering processes that lead to hot-carrier multiplication (hot-CM) \cite{Tielrooij2013, Song2013, Johannsen2013, Jensen2014, Johannsen2015, Wu2016}, where the hot-carrier density (carriers with energy above the chemical potential) exceeds the carrier density immediately after photoexcitation. Important questions concerning the ultrafast carrier dynamics, which have not been addressed experimentally, pertain to how carrier heating processes change with Fermi energy, what the different conditions are for the occurrence of CM and hot-CM, and how the ultrafast carrier dynamics correlate with the THz photoconductivity.
\

The understanding of the conductivity of unexcited graphene comes mainly from transport measurements and theoretical modeling of momentum scattering processes. Typically, in graphene devices with a mobility below 10,000 cm$^2$/Vs, i.e.\ most substrate-supported graphene devices, the conductivity is limited by long-range scattering, including for example scattering between carriers and Coulomb impurities in the substrate \cite{Tan2007, Hwang2008, DasSarma2011} or at corrugations \cite{Gibertini2012}. Only in the case of ultraclean, suspended or hBN-encapsulated graphene, much higher mobilities are achieved, because impurity scattering is suppressed. In that case, phonon scattering has to be taken into account as well to explain the observed conductivity \cite{Wang2013}. The conductivity of graphene after photoexcitation has been mainly addressed by optical pump - THz probe measurements \cite{George2008, Strait2011, Boubanga2012, Docherty2012, Tielrooij2013, Frenzel2013, Jnawali2013, Shi2014, Frenzel2014, Jensen2014, Hafez2015, Mihnev2016, Wang2017}. This technique consists of photoexcitation using an optical pump pulse that excites one electron per absorbed photon from valence to conduction band, followed by measurement of the transmission of a THz probe pulse. Given the THz photon energy of a few meV, smaller than the Fermi energy of most graphene samples, THz absorption takes place due to intraband electronic transitions and therefore provides a direct, contact-free measurement of graphene's low frequency (nearly DC) conductivity. The change in conductivity at THz frequencies following photoexcitation is referred to as the THz photoconductivity. Recent experiments on the (Fermi-energy dependent) THz photoconductivity have indicated that the sign of the photoconductivity can be both positive and negative. Explanations of this surprising effect include the modified carrier distribution due to carrier heating \cite{George2008, Strait2011, Tielrooij2013, Jensen2014}, increased momentum scattering with optical phonons \cite{Jnawali2013, Mihnev2016}, THz emission \cite{Docherty2012} and different combinations of these effects \cite{Boubanga2012, Frenzel2013, Shi2014, Frenzel2014}. Importantly, although all these studies have been performed on substrate-supported graphene devices, the effect of photoexcitation on the dominant momentum scattering mechanism -- long-range Coulomb impurity scattering -- has so far not been studied.
\

Here, we address a number of key issues in the observed ultrafast carrier dynamics and photoconductivity of graphene, for which so far controversial and partially contradicting explanations have been put forward. We investigate these issues by measuring the THz photoconductivity in a single, well-characterized, substrate-supported graphene device with tunable carrier density. Moreover, we vary the photon energy of the pump pulse, arguably the most revealing parameter for understanding photodynamical processes, such as carrier heating \cite{Tielrooij2013, Tielrooij2015} and (hot-)carrier multiplication processes \cite{Schaller2004, Pijpers2009}. We complement the experimental results with theoretical calculations that take into account all relevant energy relaxation processes, and consider that the conductivity is governed by momentum scattering through long-range interactions with impurities. This accurately reproduces the observed experimental trends without freely adjustable parameters. Our results lead to an intuitive physical picture based on two distinct regimes of carrier heating. \textit{(i)} In the case of low carrier density ($E_{\rm F} \lesssim$ 0.1 eV), energy relaxation of the initially excited electron-hole pairs occurs through electronic transitions from valence to conduction band, leading to a broadened carrier distribution, i.e.\ \textit{interband heating}. Under certain conditions, this interband heating corresponds to the creation of secondary carriers in the conduction band and thus CM, although this process competes with relaxation of the photoexcited carriers from conduction to valence band due to phonon emission and other carrier-carrier scattering channels. Here, photoexcitation and the subsequent ultrafast dynamics lead to an increase in THz conductivity -- positive THz photoconductivity, which can be partially understood from the additional free carriers that are created. \textit{(ii)}  For higher initial carrier densities ($E_{\rm F} \gtrsim$ 0.15 eV), interband heating is suppressed. The most effective ultrafast energy relaxation channel consists of electronic transitions from below the chemical potential to above the chemical potential, i.e.\ \textit{intraband heating}. Under the conditions studied here, this intraband heating corresponds to the creation of hot carriers in the conduction band and thus hot-CM. Here, photoexcitation and the subsequent ultrafast dynamics lead to a reduction of the conductivity -- negative THz photoconductivity -- mainly due to reduced screening of the impurities, leading to increased momentum scattering. In short, we find that the transition from interband to intraband heating and the transition from positive to negative THz photoconductivity are determined by the ultrafast carrier dynamics following photoexcitation, and are thus naturally correlated. Finally, we show that energy is efficiently transduced from incident light to graphene carriers, regardless of the Fermi energy, meaning that both interband and intraband heating are efficient. Thus, our results show that a large collection of experimental observations related to photoexcited graphene can be explained by a simple and intuitive picture, where energy relaxation occurs through ultrafast and efficient carrier heating, and the photoconductivity is the result of modified screening of long-range Coulomb scattering.  
\

\begin{figure}[h!]
\centering
\includegraphics[scale=0.7]{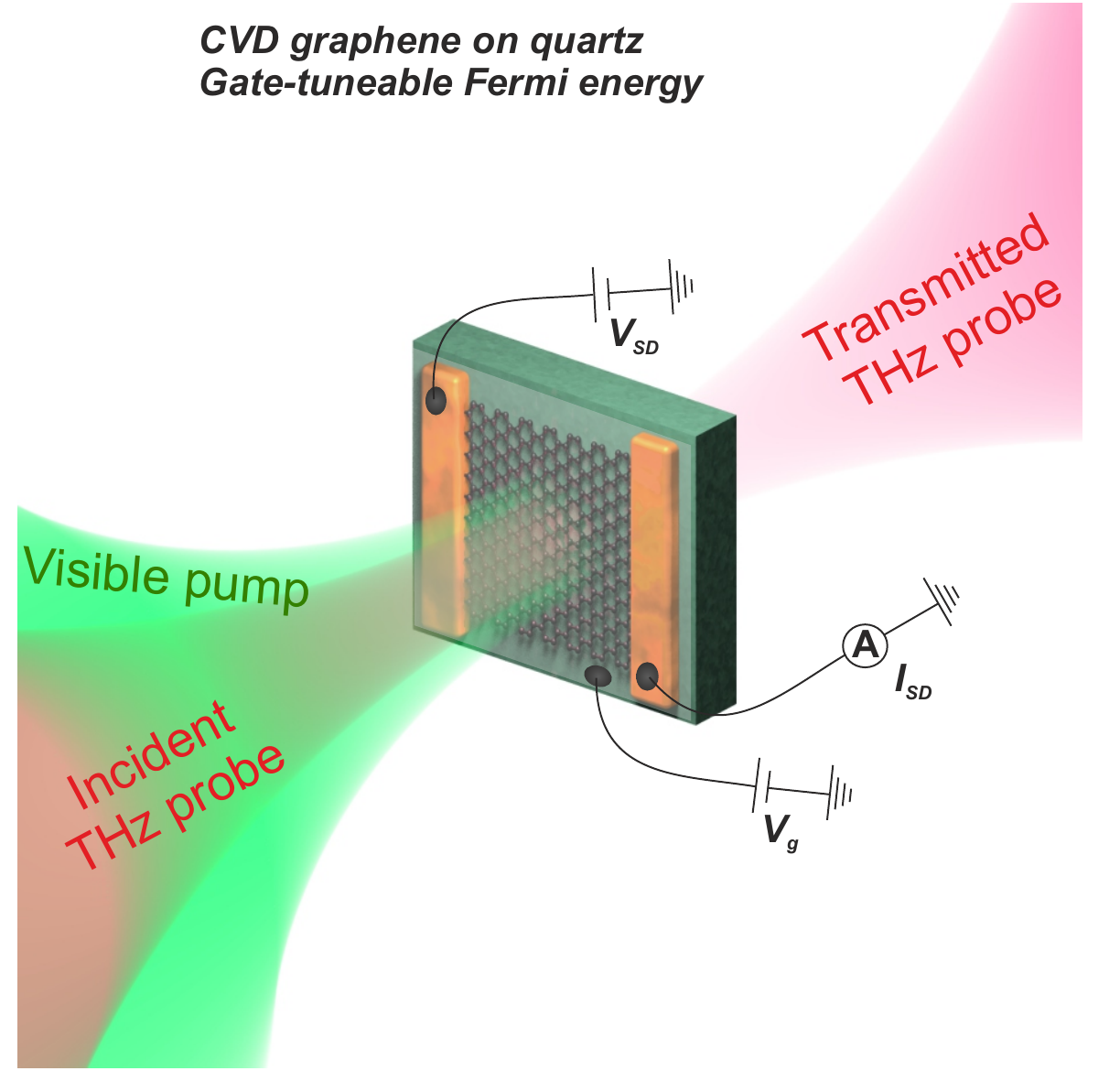}
\caption{Illustration of the optical pump -- THz probe measurement technique and the gate-tunable graphene device design, as explained in the main text and Methods. By applying a gate voltage $V_{\rm g}$ to the polymer electrolyte, we change the Fermi energy of graphene and thereby its DC conductivity $\sigma_0$. By applying a source-drain voltage $V_{\rm SD}$ between the two contacts and measuring the current $I_{\rm SD}$ we extract $\sigma_0$ (after correcting for contact resistance).} \vspace{12pt}
\label{Fig1}
\end{figure}

\section*{Results}

\subsubsection*{Photoconductivity vs. carrier density}

\begin{figure}[h!]
\centering
\includegraphics[scale=0.84]{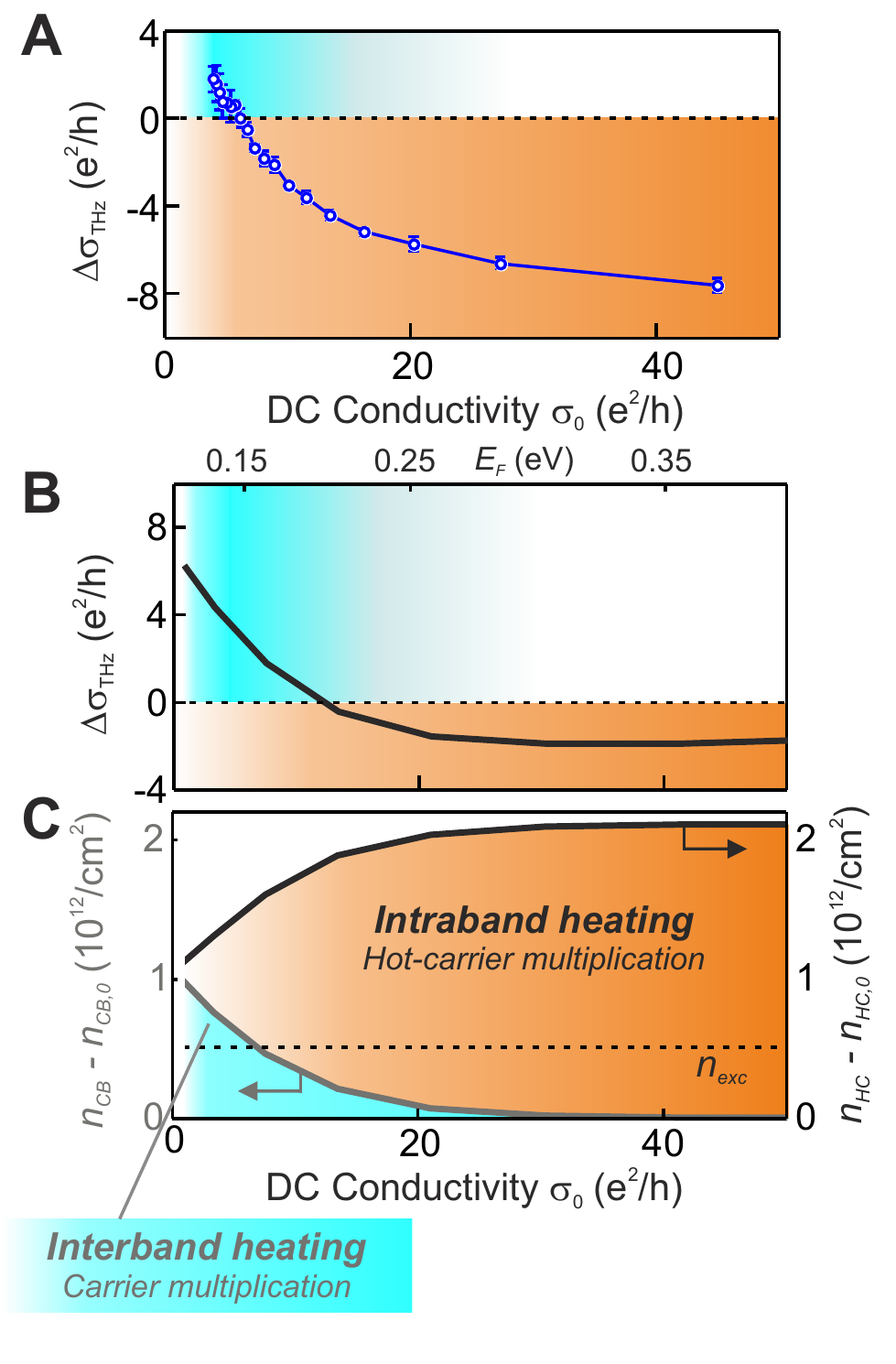}
\caption{\textbf{A)} The measured THz photoconductivity $\Delta \sigma_{\rm THz}$ as a function of simultaneously measured DC conductivity $\sigma_{\rm 0}$, following a pump pulse with a wavelength of 800 nm that creates a photoexcited carrier density $n_{\rm exc} = $ 0.5$\times 10^{12}$/cm$^{2}$. \textbf{B)} Theoretical calculation (see Methods) of the THz photoconductivity vs. DC conductivity (bottom horizontal axis) or Fermi energy (top horizontal axis) for the same parameters as in the experiment, at time $t =$ 300 fs after photoexcitation. The theoretical results (which are free of adjustable parameters) reproduce the magnitude of the signal within a factor $\lesssim2$ and the positive-to-negative transition. \textbf{C)} The calculated carrier density in the conduction band (left vertical axis) and hot-carrier density (right vertical axis) with respect to equilibrium, for the same parameters as in the experiment. The dashed horizontal line indicates $n_{\rm exc}$, which is the same for all equilibrium DC conductivities. Interband heating occurs when $n_{\rm CB} - n_{\rm CB,0} >$ 0, whereas intraband heating occurs when $n_{\rm HC} - n_{\rm HC,0} >$ 0.}\vspace{12pt}
\label{Fig2}
\end{figure}

We use optical pump -- THz probe spectroscopy as illustrated in \textcolor{blue}{\textbf{\ref{Fig1}}}. A visible pump pulse with a duration of $\sim$100 fs, spot size of $\sim$1 cm and a photon energy $E_{\rm ph}$ typically between 1.0 and 2.5 eV excites electrons from valence band states with energy $-E_{\rm ph}/2$ to conduction band states with energy $E_{\rm ph}/2$. After a time delay $t\sim$ 300 fs (see \textcolor{blue}{\textbf{\ref{Fig9}A}}) we probe the pump-induced change in the transmission of a quasi-single cycle THz pulse (center frequency $\sim$0.6 THz). This yields the time-dependent THz photoconductivity $\Delta \sigma_{\rm THz} = \sigma_{\rm THz}(t) - \sigma_{\rm THz}(0)$, where $\sigma_{\rm THz}(0)$ is the THz conductivity at equilibrium, i.e.\ before the pump pulse. The device in \textcolor{blue}{\textbf{\ref{Fig1}}} (see Methods for details) consists of CVD graphene on quartz with a drop cast ionic gate deposited on top to control the  equilibrium carrier density. Through resistance measurements using the source and drain contacts we extract the equilibrium DC conductivity $\sigma_{\rm 0}$ (see also \textcolor{blue}{\textbf{\ref{Fig9}B}}). We infer a mobility of $\sim$1000 cm$^2$/Vs by combining the transport measurements with Raman measurements to determine the carrier density \cite{Das2008} (see \textcolor{blue}{\textbf{\ref{Fig9}C}}). This mobility confirms that the conductivity is indeed governed by impurity scattering \cite{Tan2007, Hwang2008, DasSarma2011}. It also indicates a momentum scattering time on the order of tens of femtoseconds, meaning that the THz (photo)conductivity is mainly real and has very little frequency dependence in our experimental THz window ($\sim$0.4 -- 1.8 THz). \textcolor{blue}{\textbf{\ref{Fig2}A}} shows the optically measured THz photoconductivity $\Delta \sigma_{\rm THz}$ as a function of the equilibrium DC conductivity $\sigma_{\rm 0}$, which is obtained simultaneously using electrical measurements (see Methods for details). We used a fixed fluence of 8 $\mu$J/cm$^2$ and a pump wavelength of 800 nm, and observe a transition from a positive signal for low steady-state conductivity (low carrier density) to a negative signal for higher steady-state conductivity (higher carrier density). A similar sign change of the THz photoconductivity was observed in Refs.\ \cite{Frenzel2014, Shi2014, Jensen2014, Wang2017}.
\

To understand the physics behind these experimental results, we show in \textcolor{blue}{\textbf{\ref{Fig2}B}} the results of our theoretical calculation of the THz photoconductivity. To model the system, we first solve a semi-classical Boltzmann equation that takes into account carrier-carrier and carrier-optical phonon scattering in graphene. Our method has been described in detail in Refs.\ \cite{Brida2013, Tomadin2013}. The calculation is non-perturbative and yields the full non-equilibrium electron and optical phonon distributions in time following photoexcitation. We do not integrate the Boltzmann equation during the photoexcitation process, but use as initial condition for the electron distribution directly following photoexcitation, a Fermi-Dirac distribution with an additional Gaussian population of photoexcited electrons in the conduction band (centered around $E_{\rm ph }/2$) and a corresponding population of holes in the valence band (centered around $-E_{\rm ph }/2$). For the sake of comparison with experiment, the Fermi-Dirac distribution always has a finite carrier density in the conduction band, i.e.\ positive Fermi energy. We verify that, after a short initial transient $t \lesssim$ 20 fs, the quasi-equilibrium electron distribution consists of two Fermi-Dirac distributions, in valence and conduction bands, with the same temperature but different chemical potentials. The values of the temperature and the chemical potentials, obtained by fitting the electron distribution, are then used to calculate the photoconductivity using the semiclassical Boltzmann equation linearized around the quasi-equilibrium distribution, taking into account scattering between carriers and long-range Coulomb impurities (see Methods for details). The measured device mobility of $\sim$1000 cm$^2$/Vs dictates a surface impurity density of 5$\times$10$^{12}$/cm$^2$. 
\

In the following paragraphs, we will present experimental and theoretical results that elucidate the processes that dominate the ultrafast dynamics following photoexcitation of graphene. Subsequently, we will discuss in detail the reason for the transition from positive to negative photoconductivity. First, we point out that the results in \textcolor{blue}{\textbf{\ref{Fig2}B}} are not obtained by fitting our theoretical calculation to the experimental results in \textcolor{blue}{\textbf{\ref{Fig2}A}}, so the agreement between the two is remarkable. Our calculation of the conductivity uses the general Boltzmann equation, rather than simplified versions, such as the Drude equation with separate contributions from momentum scattering and from changes in the electron distribution, captured by the Drude weight \cite{Jnawali2013, Frenzel2013, Frenzel2014, Shi2014}, or the Sommerfeld expansion of the Boltzmann equation, which is only valid for high doping and low fluence, such that the electron temperature is lower than the Fermi temperature $T_{\rm F} = E_{\rm F}/k_{\rm B}$, with $k_{\rm B}$ the Boltzmann constant \cite{Tielrooij2013}. Furthermore,  we only consider the dominant momentum scattering process, namely long-range impurities scattering, rather than assuming a modified momentum scattering rate due to enhanced phonon interactions, as in Refs.\ \cite{Frenzel2014, Shi2014, Hafez2015, Mihnev2016}. Including this secondary effect would likely improve the quantitative agreement between the experimental and the theoretical results. Further improvement could come from including effects of electron-hole puddles, which could be substantial in substrate-supported CVD graphene devices. Finally, we point out that the photoconductivity is a {\it non-equilibrium} property and, as such, its magnitude and sign change with time in a fashion that depends on the initial conditions (e.g.\ the equilibrium carrier density) and on the details of the perturbation, such as the fluence (see e.g. Ref.\ \cite{Jensen2014}) and the frequency of the pump pulse.

\subsubsection*{Interband-to-intraband heating transition}

The calculation of the evolution of the electron distribution with time allows us to get insight into the ultrafast energy relaxation that takes place after photoexcitation. To this end, we calculate the carrier density in the conduction band before ($n_{\rm CB,0}$) and after ($n_{\rm CB}$) photoexcitation, according to $n_{\rm CB} = \int_{0}^{\infty} d\varepsilon \nu(\varepsilon) f_{0}(\varepsilon; t)$, where $\varepsilon$ is the carrier energy and $\nu(\varepsilon)$, $f_{0}(\varepsilon; t)$ are defined in the Methods section. We also calculate the hot-carrier density, i.e.\ the density of electrons in the conduction band with an energy exceeding the chemical potential $\mu_{\rm e}$, before ($n_{\rm HC,0}$) and after ($n_{\rm HC}$) photoexcitation, according to $n_{\rm HC} = \int_{\mu_{\rm e}(t)}^{\infty} d\varepsilon \nu(\varepsilon) f_{0}(\varepsilon; t)$, where $\mu_{\rm e}(t)$ is defined in the Methods section. These quantities are shown in \textcolor{blue}{\textbf{\ref{Fig2}C}} as functions of the equilibrium DC conductivity. We find that at low equilibrium DC conductivity, i.e.\ close to the charge neutrality point, photoexcitation leads to an increase in the carrier density in the conduction band: $n_{\rm CB} - n_{\rm CB,0} >$ 0. This means that \textit{interband} scattering processes, where electrons/holes are transferred between conduction/valence band, have occurred. At high conductivity, i.e.\ away from the charge neutrality point, photoexcitation does not lead to a change in the carrier density in the conduction band ($n_{\rm CB} \approx n_{\rm CB,0}$), whereas it leads to an increase in the hot-carrier density: $n_{\rm HC} - n_{\rm HC,0} >$ 0. This means that \textit{intraband} scattering events have occurred. Obviously, upon approaching the Dirac point, the hot carrier density converges to the conduction band carrier density, as we see in \textcolor{blue}{\textbf{\ref{Fig2}C}}.
\

\begin{figure}[h!]
\centering
\includegraphics[scale=0.62]{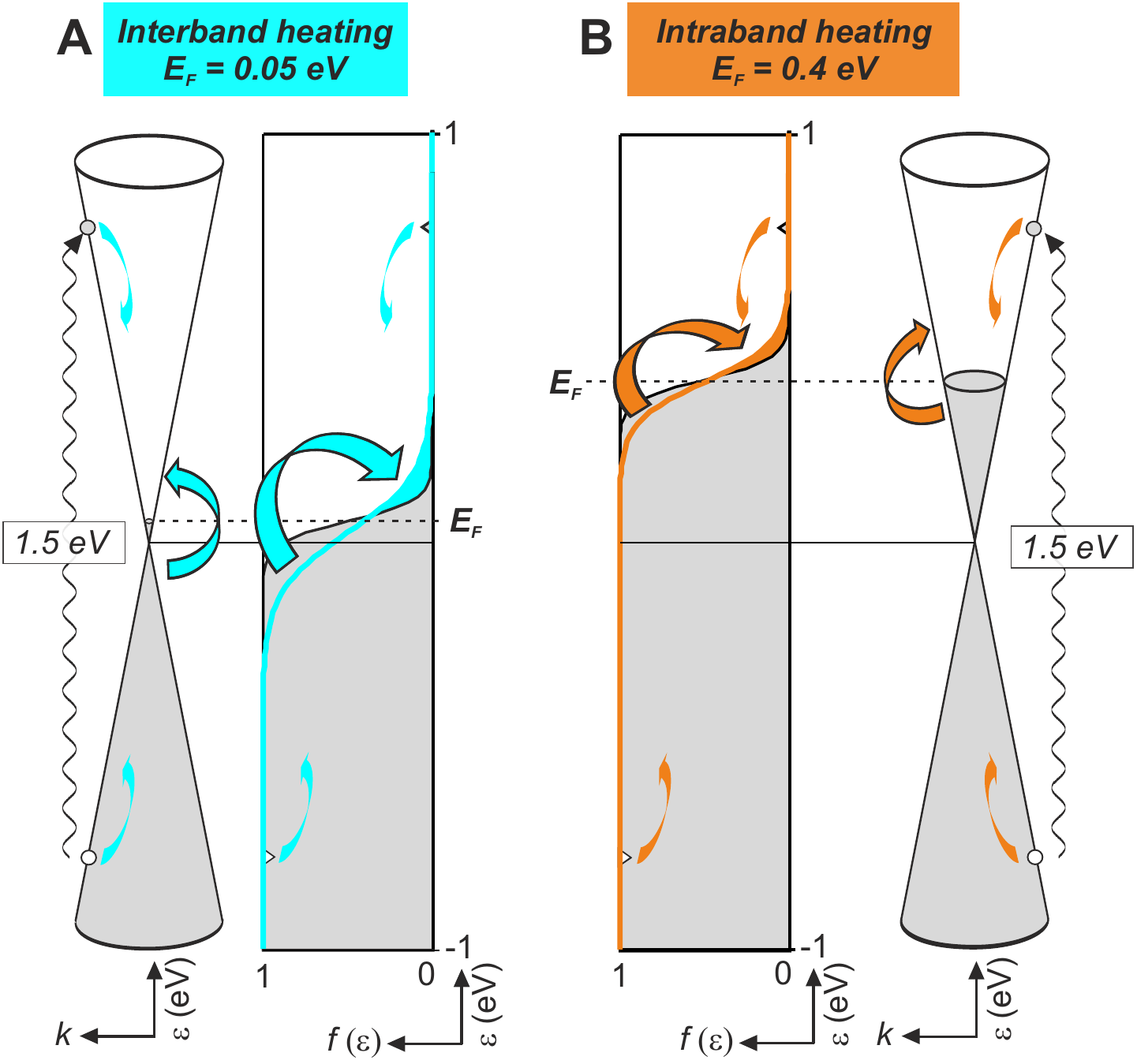}
\caption{
The panels show the calculated electron distribution $f(\varepsilon)$ as a function of energy $\varepsilon$, at time $t = $ 0 before photoexcitation (black lines and gray areas) and at a time $t = $ 300 fs (cyan and orange lines and areas). The state occupation in the Dirac cones is also depicted alongside the electron distribution. The wavy lines represent the photoexcitation of electrons (holes) in conduction (valence) band at energy $E_{\rm ph} / 2$ ($-E_{\rm ph} / 2$). The pump photons have energy $E_{\rm ph} = $ 1.5 eV.
\textbf{A)} At the equilibrium Fermi energy $E_{\rm F} \simeq$ 0.05 eV, the broadening of the electron distribution involves electronic transitions from valence to conduction band, represented by cyan arrows, i.e.\ interband heating.
\textbf{B)} At $E_{\rm F} \simeq $ 0.4 eV, the broadening is mostly due to electronic transitions from below to above the chemical potential, represented by orange arrows, i.e.\ intraband heating. } \vspace{12pt}
\label{Fig3}
\end{figure}

The physical interpretation of this interband-to-intraband scattering transition becomes clear when we examine the electron distribution before and after photoexcitation. We calculate these for an equilibrium Fermi energy of 0.05 eV (\textcolor{blue}{\textbf{\ref{Fig3}A}}) and 0.4 eV (\textcolor{blue}{\textbf{\ref{Fig3}B}}), corresponding to $\sigma_{0} \lesssim$ 5 $e^2$/$h$ and $\sigma_{0} \simeq$ 50 $e^2$/$h$, respectively. Before photoexcitation, at room temperature $T = $ 300 K, the distributions are slightly broadened and, for both equilibrium Fermi energies, the hole density in the valence band is small compared to the electron density in the conduction band. At time $t = $ 300 fs following photoexcitation with $E_{\rm ph} =$ 1.5 eV, the electron distributions for the two equilibrium Fermi energies have broadened substantially, indicating carrier heating. However, the broadened distributions are qualitatively different, owing to different relaxation processes dominating the dynamics depending on the equilibrium Fermi energy. At $E_{\rm F} \simeq$ 0.05 eV, the heated electron distribution contains a significant hole density in the valence band and an increased electron density in the conduction band (see cyan-shaded area in \textcolor{blue}{\textbf{\ref{Fig3}A}}). This indicates that electronic transitions have taken place from valence band to conduction band. In contrast, at $E_{\rm F} \simeq$ 0.4 eV, the heated electron distribution does not contain holes in the valence band. However, it contains an increased hot electron density (see orange-shaded area in \textcolor{blue}{\textbf{\ref{Fig3}B}}). Thus, we see that the occurrence of both interband and intraband transitions can be understood from heating of the graphene carriers. Furthermore, we see that the transition from positive to negative photoconductivity correlates with a transition from \textit{interband} heating to dominant \textit{intraband} heating. 
\

Since our calculations show an increased (hot) carrier density for low (high) equilibrium Fermi energy, this prompts us to investigate if (hot-)carrier multiplication occurs. We study this by comparing the electron distributions before and after photoexcitation to the absorbed photon density, which equals the initial density of photoexcited carriers $n_{\rm exc }$. CM takes place when the carrier density in the conduction band with respect to equilibrium, $n_{\rm CB} - n_{\rm CB,0}$, is larger than $n_{\rm exc}$. Hot-CM takes place when the hot-carrier density with respect to equilibrium, $n_{\rm HC} - n_{\rm HC,0}$, is larger than $n_{\rm exc}$. Our calculations show that CM can take place for low equilibrium DC conductivity, i.e.\ close to the charge neutrality point (see \textcolor{blue}{\textbf{\ref{Fig2}C}}). Upon increasing the carrier density, CM disappears as a result of suppressed interband heating, whereas hot-CM increases with increasing carrier density, as result of intraband heating.

\begin{figure}[h!]
\centering
\includegraphics[scale=0.8]{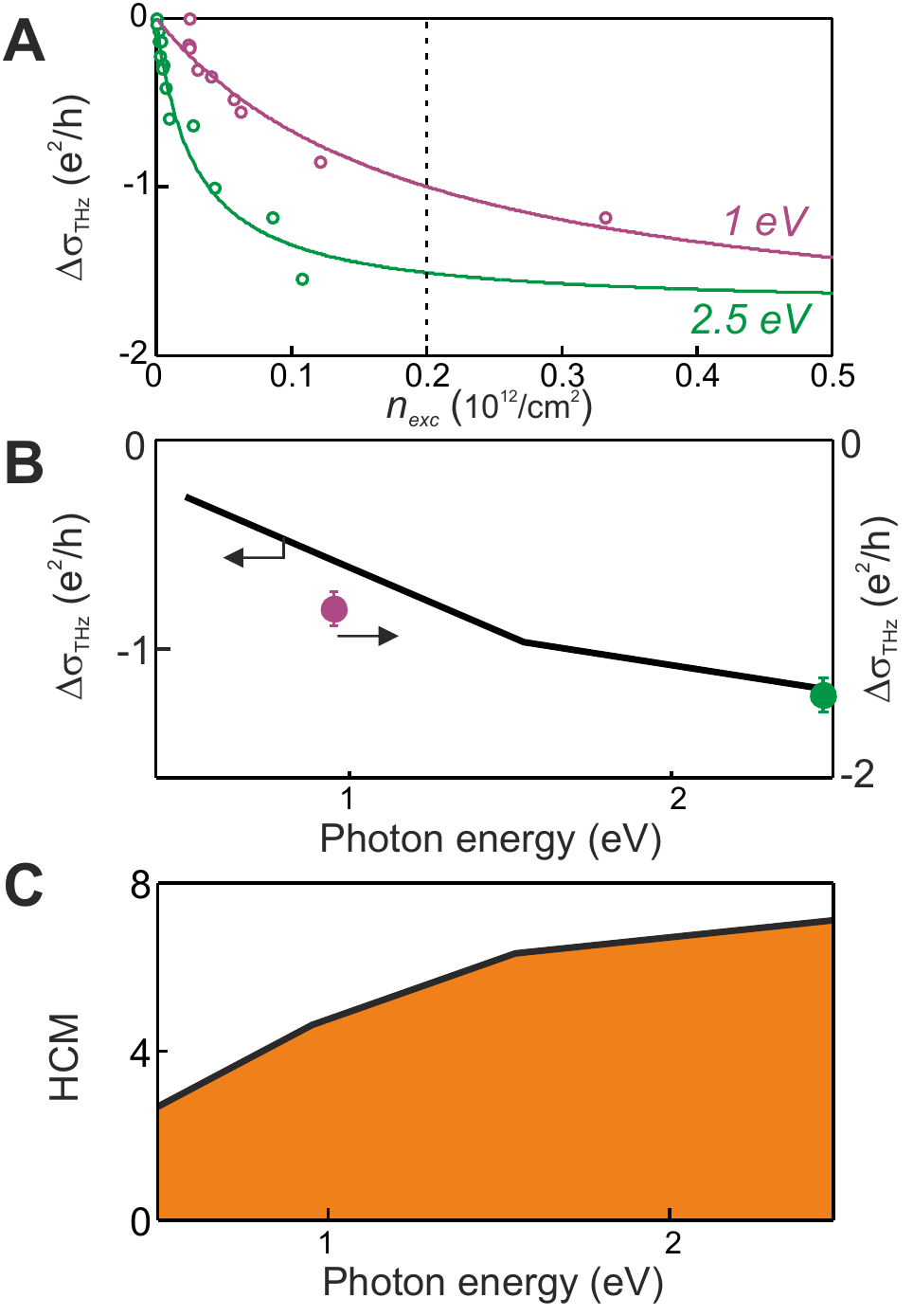}
\caption{
\textbf{A)} The THz photoconductivity $\Delta \sigma_{\rm THz}$ as a function of pump pulse fluence, parametrized by the photoexcited carrier density $n_{\rm exc}$, for two pump photon energies $E_{\rm ph} = $ 1.0 (purple circles) and 2.5 eV (green circles). The solid lines serve as guides to the eye based on saturation curves. These measurements are done at $V_{\rm g}$ = 0 V, where we measure a DC conductivity of $\sigma_{\rm 0} \approx$ 8 e$^2$/h. Using the mobility of $\sim$1000 cm$^2$/Vs, we extract an equilibrium Fermi energy of $\lvert E_{\rm F}\rvert \simeq$ 0.25 eV, corresponding to graphene gated away from the charge neutrality point. The THz photoconductivity saturates for $n_{\rm exc} \gtrsim 0.05 \times 10^{12}$/cm$^2$, corresponding to a relatively low incident fluence $<1$ $\mu J$/cm$^2$, as observed earlier \cite{Jnawali2013, Jensen2014}.
The signal (and therefore the hot carrier density) is larger for the larger photon energy. Thus, larger photon energies are responsible for larger hot carrier densities. This is a key signature of efficient intraband heating \cite{Tielrooij2013}.
\textbf{B)} The THz photoconductivity at $n_{\rm exc} = 0.2 \times 10^{12}$/cm$^2$ (vertical dashed line in panel \textbf{A}), as a function of the pump photon energy. The solid circles are experimental data (right vertical axis) and the solid line the result of the theoretical calculation (left vertical axis) for $E_{\rm F}$ = 0.25 eV. Experiment and theory show good qualitative agreement, with a larger absolute value of the THz photoconductivity for larger photon energy.
\textbf{C)} The calculated hot-carrier multiplication factor HCM, defined in the main text, as a function of the pump photon energy. The HCM-factor increases with photon energy. } \vspace{12pt}
\label{Fig4}
\end{figure}

\subsubsection*{Photoconductivity vs. photon energy}

\begin{figure}[h!]
\centering
\includegraphics[scale=0.8]{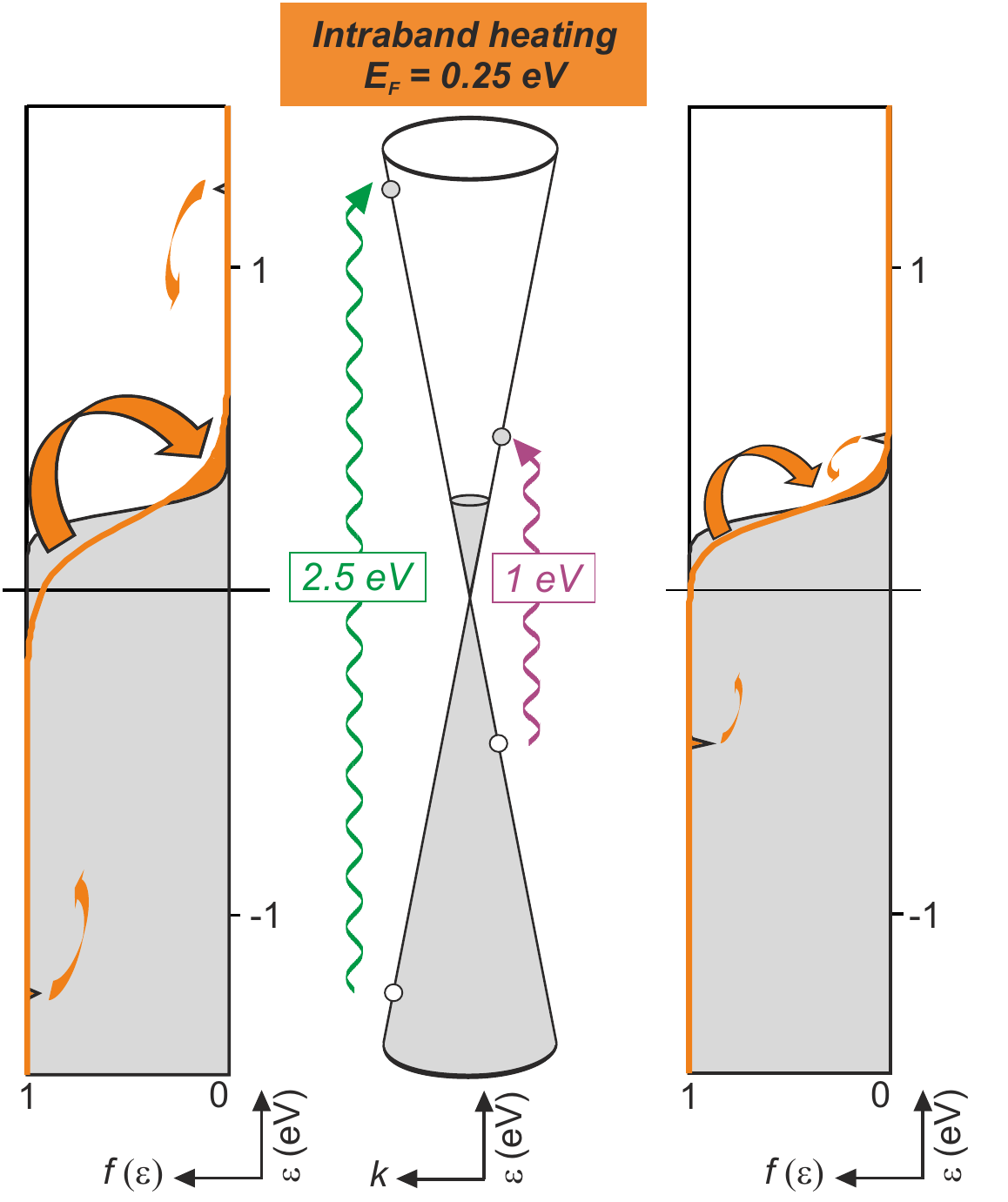}
\caption{
The electron distributions and the state occupation in the Dirac cone, cfr.\ Fig.\ 3, before and after photoexcitation with pump photon energy $E_{\rm ph} = $ 2.5 eV (left, green wavy line) and 1.0 eV (right, purple wavy line). The broadening of the electron distribution is larger in the $E_{\rm ph} = $ 2.5 eV case, indicating more intraband heating for larger photon energy.} \vspace{12pt}
\label{Fig5}
\end{figure}

\begin{figure}[h!]
\centering
\includegraphics[scale=0.8]{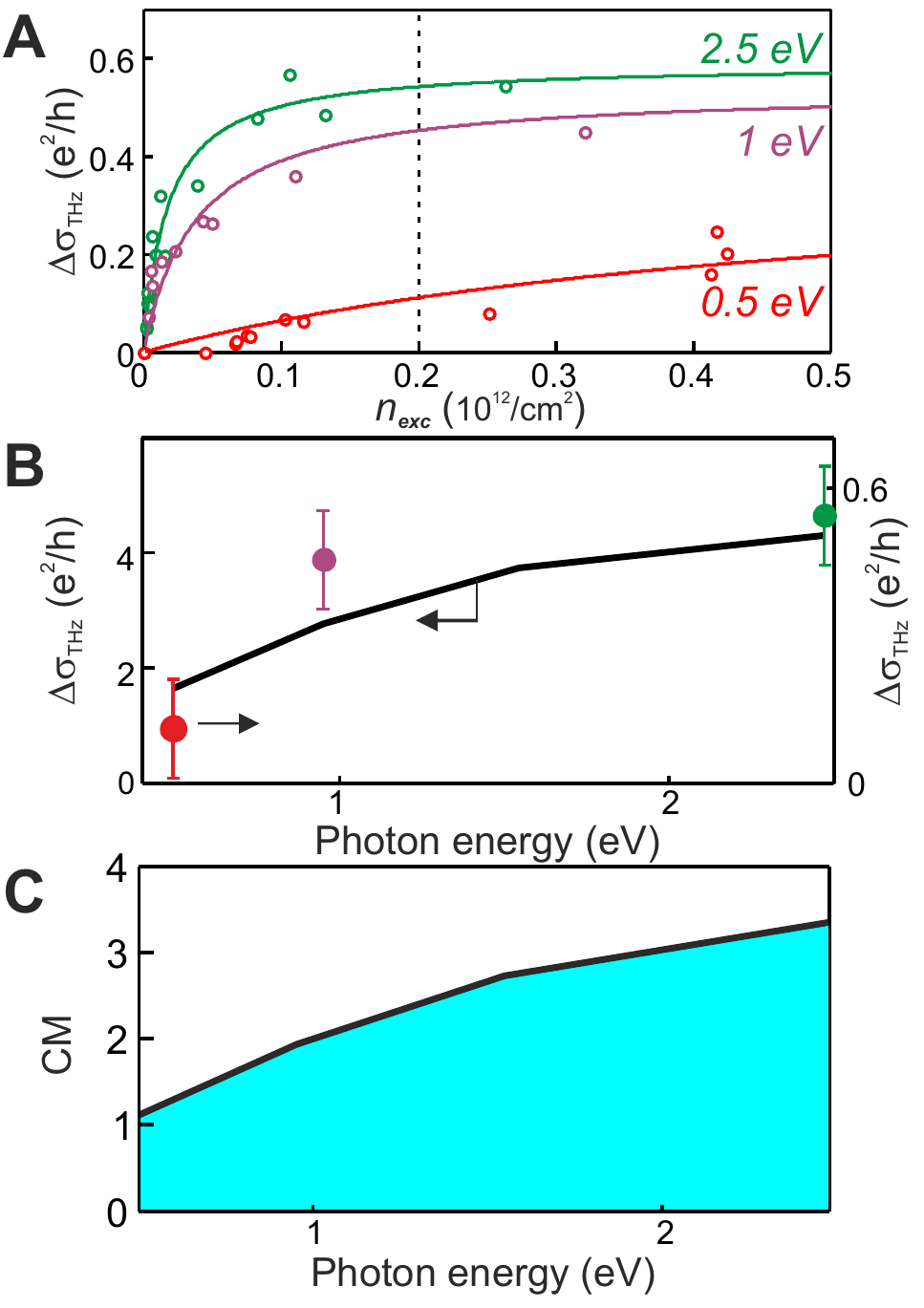}
\caption{This figure is analogous to Fig.\ 4, with the difference that the measurements are done at $V_{\rm g}$ = 0.5 V, the point of lowest DC conductivity ($\sigma_{\rm 0} \approx$ 2 e$^2$/h), corresponding to graphene close to the charge neutrality point. The calculations are done for $E_{\rm F}$ = 0.05 eV, although due to puddles the effective Fermi energy in the experiment could be significantly larger. We observe that the value of the THz photoconductivity is positive, while in Fig.\ 4 it is negative. Panels \textbf{A)} and \textbf{B)} show the THz photoconductivity for three pump photon energies $E_{\rm ph} = $ 0.5, 1.0, and 2.5 eV. Panel \textbf{C)} shows the calculated carrier multiplication factor CM, defined in the main text. To understand the magnitude of the photoconductivity, we note that the added carrier density in the conduction band corresponds to $\Delta n$ = CM$\cdot n_{\rm exc}$. Using CM = 3 (for 2 eV photon energy), $n_{\rm exc}$ = 0.2$\cdot$10$^{12}$/cm$^2$, and a mobility of $\mu$ = 1000 cm$^2$/Vs, we obtain an increased conductivity of $\Delta \sigma = \Delta n e\mu$ = 4 e$^2$/h. This is agrees with the calculated increase in conductivity (see Fig.\ 6B for 2 eV photon energy). The measured value is significantly lower, most likely due to electron-hole puddles.} \vspace{12pt}
\label{Fig6}
\end{figure}

\begin{figure}[h!]
\centering
\includegraphics[scale=0.8]{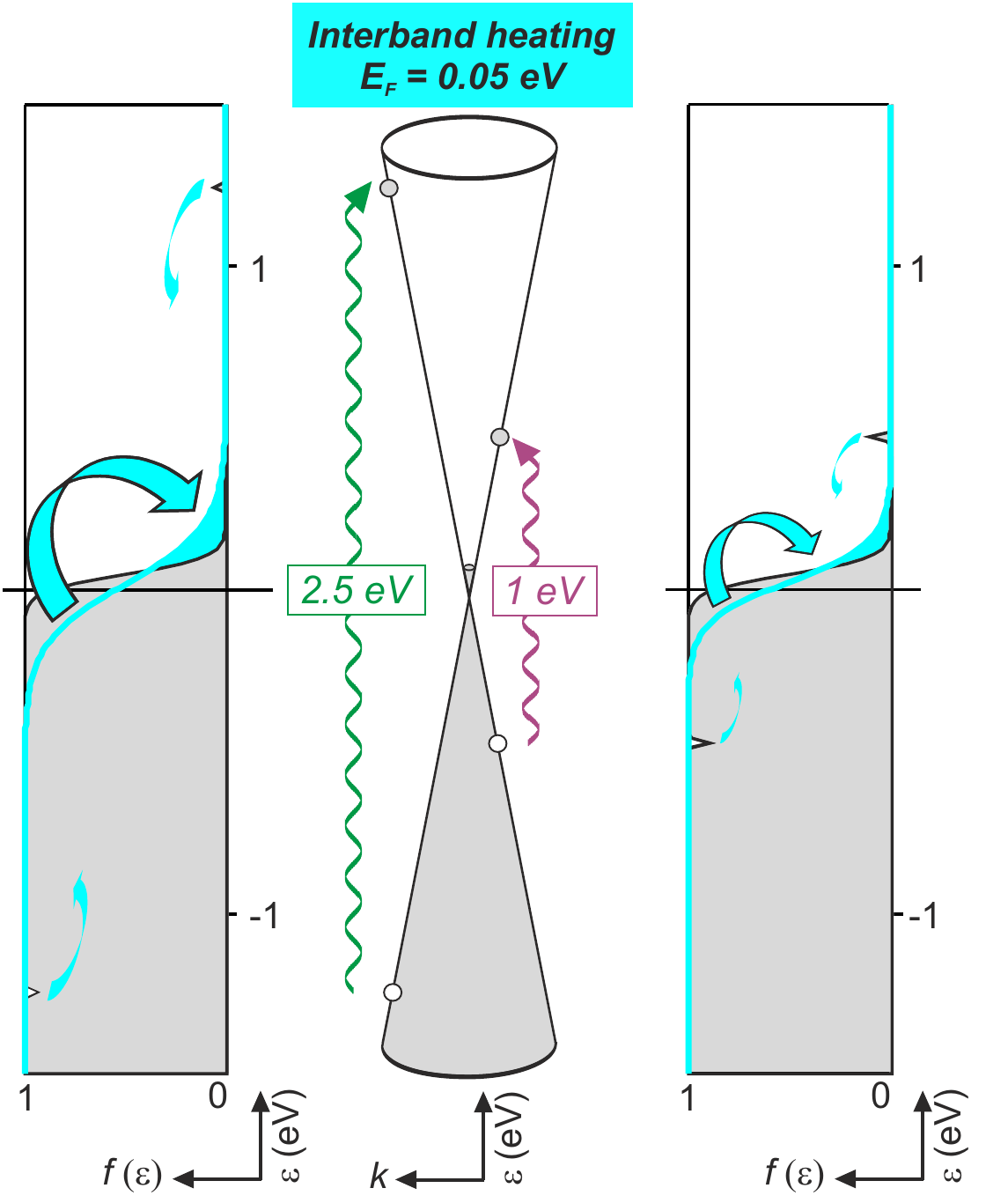}
\caption{This figure shows that interband heating contributes to the broadening of the electron distribution, and that the broadening of the electron distribution is larger in the $E_{\rm ph} = $ 2.5 eV case, compared to the 1 eV case, indicating more interband heating for larger photon energy.} \vspace{12pt}
\label{Fig7}
\end{figure}

Our optical -- pump THz probe experiment (and the same holds for other all-optical experiments) does not give direct insight into (hot-)carrier multiplication numbers, as we do not harvest electrons through an electrical measurement, which would allow for counting electrons and comparing with absorbed photons. However, a key indication of the occurrence of efficient carrier heating, as well as (hot-)CM, is that a larger photon energy leads to a larger density of secondary carriers (hot carriers) \cite{Tielrooij2013, Song2013, Tielrooij2015}. The reason for this is that additional energy from photons is transduced to additional excited carriers in the electronic system. Figures 4--7 explore this trend in our experiments. The results in \textcolor{blue}{\textbf{\ref{Fig4}}} and \textcolor{blue}{\textbf{\ref{Fig5}}} correspond to the device tuned away from the charge neutrality point, with an estimated Fermi energy $\lvert E_{\rm F}\rvert \simeq$ 0.25 eV (the ionic gate is at 0 V, where we measure $\sigma_{\rm 0} \approx$ 8 e$^2$/h). Here, \textit{interband} heating is suppressed, whereas \textit{intraband} heating takes place. \textcolor{blue}{\textbf{\ref{Fig6}}} and \textcolor{blue}{\textbf{\ref{Fig7}}}, instead, correspond to 0.5 V on the ionic gate, where we measure $\sigma_{\rm 0} \approx$ 2 e$^2$/h. Here, the carrier density is low, and \textit{interband} heating is present.
\

\begin{figure}[hhhhhhhh!!!!!!!!]
\centering
\includegraphics[scale=0.54]{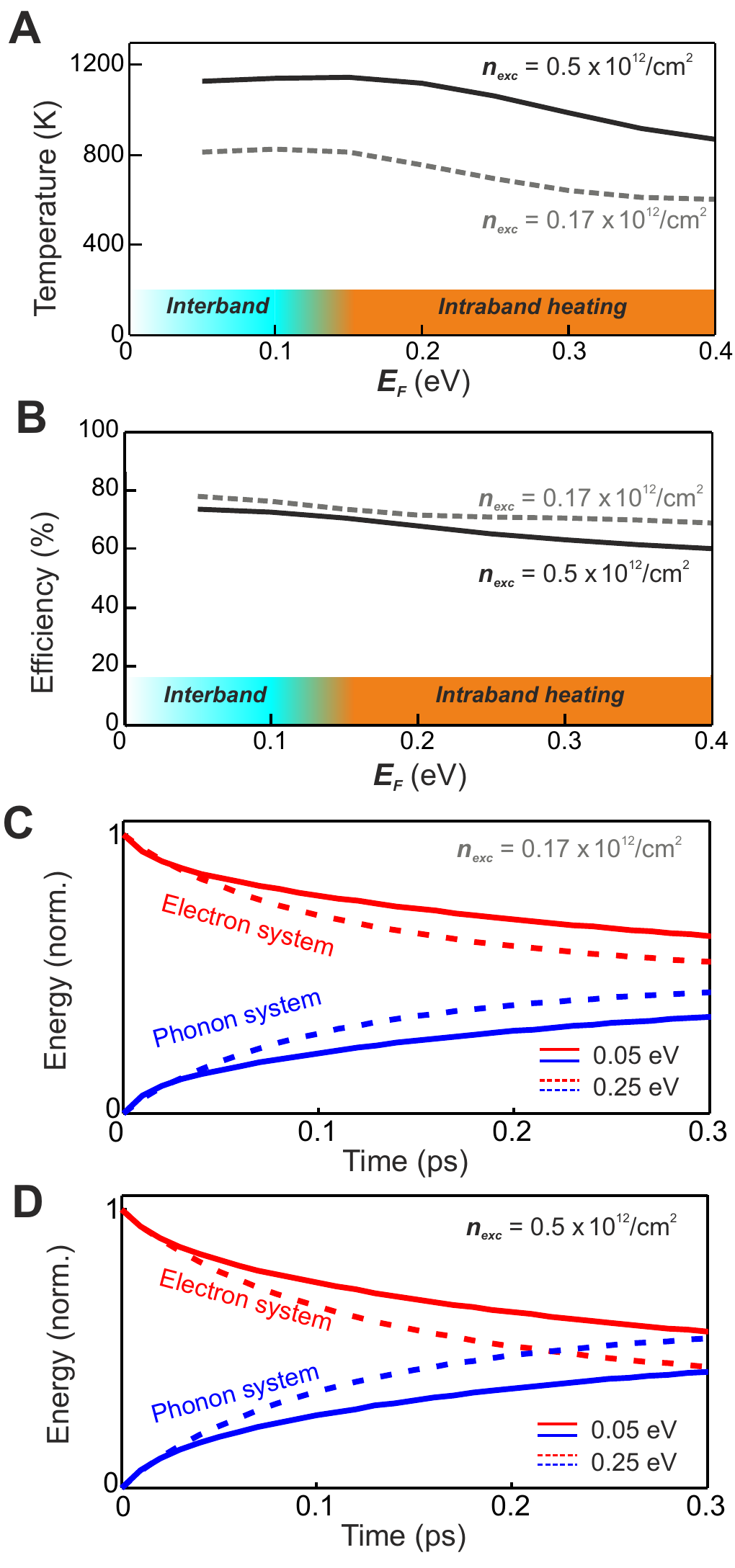}
\caption{
\textbf{A)} The electron temperature as a function of equilibrium Fermi energy at time $t = $ 100 fs after photoexcitation, for photoexcited carrier densities $n_{\rm exc} = 0.5 \times 10^{12}$/cm$^2$ (black, solid line) and $n_{\rm exc} = 0.17\times 10^{12}$/cm$^2$ (gray, dashed line). 
\textbf{B)} The energy transfer efficiency as a function of equilibrium Fermi energy at time $t = $ 100 fs after photoexcitation, for photoexcited carrier densities $n_{\rm exc} = 0.5 \times 10^{12}$/cm$^2$ (black, solid line) and $n_{\rm exc} = 0.17\times 10^{12}$/cm$^2$ (gray, dashed line). The energy transfer efficiency is defined as the ratio between the electron energy density at time $t$ and the photon energy density of the pump pulse, and is a highly relevant parameter for photodetectors based on carrier heat, among others. The equilibrium Fermi energy intervals where interband heating or intraband heating are dominant are shaded along the horizontal axes of the panels. These results show that, regardless of the occurrence of interband or intraband heating, photoexcitation leads to an increase of the electron temperature that is larger for the larger photoexcited density. The energy transfer efficiency is 60-70\% for the higher fluence and 70-80\% for the lower fluence. This shows that the energy of the absorbed pump photons is efficiently retained by the electron system and not dissipated into phonons. We also show the temporal evolution of energy dissipation in the system for low fluence \textbf{(C)} and higher fluence \textbf{(D)}. The dynamics indicate that for several hundred  femtoseconds the energy that is deposited by photons into the electronic system is still mainly present as energy in the electrons -- electron heat. }\vspace{12pt}
\label{Fig8}
\end{figure}

\begin{figure}[hhhh!!!!]
\centering
\includegraphics[scale=0.75]{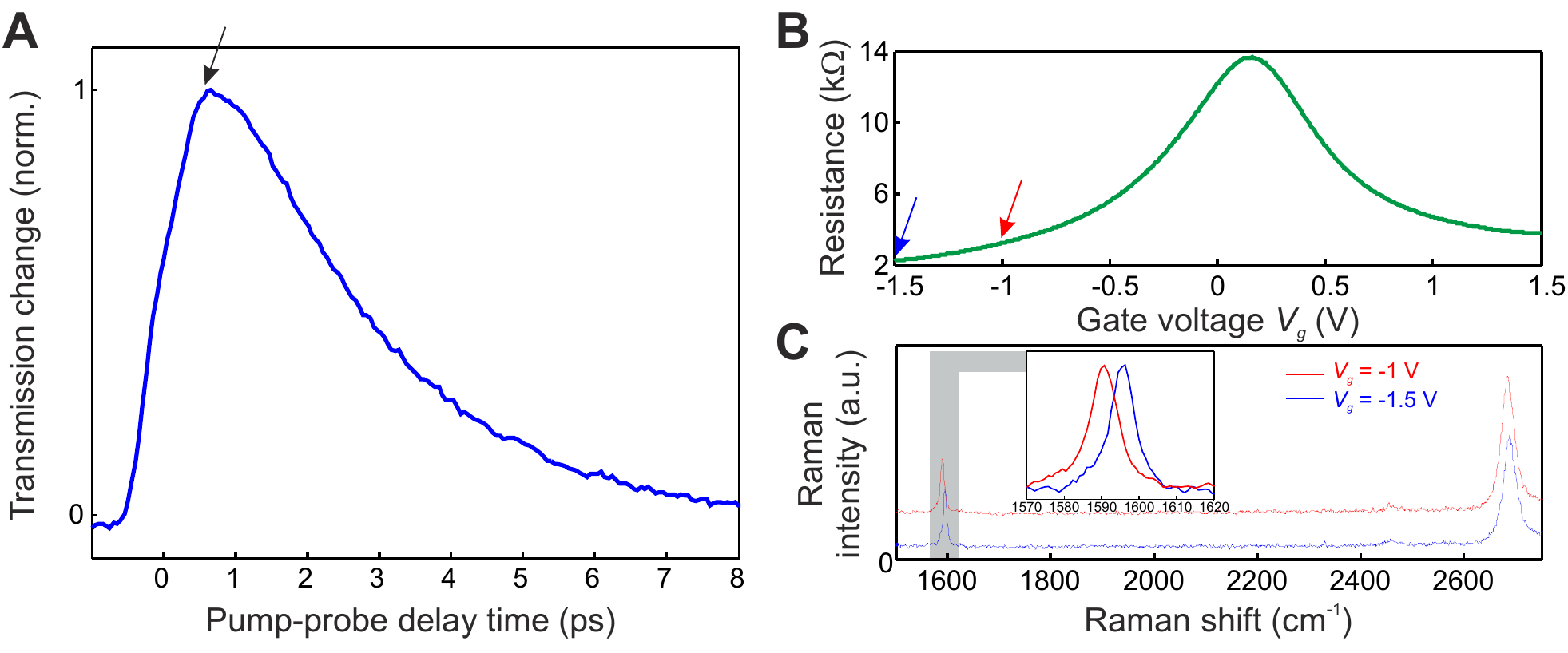}
\caption{
\textbf{A)} Normalized change in THz transmission as a function of pump-probe delay time for the gate-tunable device. The 800 nm pump pulse coincides with the peak of the THz pulse at time zero. The black arrow indicates the time delay where we typically examine the THz photoconductivity. 
\textbf{B)} Transport measurements of the device with resistance as a function of gate voltage. The Dirac point lies at slightly positive voltage, whereas changing the gate voltage clearly leads to bipolar doping behavior. The blue and red arrows indicate the gate voltages where Raman spectra are taken.  
\textbf{C)} Raman spectra of the device at $V_g = -1.5$ V ($V_g = -1$ V) in blue (red), with more detail of the carrier-density sensitive G peak in the inset. We extract a carrier density of 8$\cdot$10$^{12}$/cm$^2$ (5$\cdot$10$^{12}$/cm$^2$) using Ref.\ \cite{Das2008}. We use these extracted carrier densities together with the measured conductivities from transport in panel $\textbf{B}$ (corrected for an estimated contact resistance of 1.5 k$\Omega$) to obtain a mobility of $\sim$1000 cm$^2$/Vs. The width of the 2D peak is $\sim$25--30 cm$^{-1}$. } \vspace{12pt}
\label{Fig9}
\end{figure}

\textcolor{blue}{\textbf{\ref{Fig4}A}} shows the measured (peak) THz photoconductivity as a function of pump pulse fluence, parametrized by the photoexcited carrier density, where wavelength-dependent absorption is taken into account. The results for two pump photon energies of 1.0 eV and 2.5 eV are shown. The larger photon energy leads to a larger negative photoconductivity at constant excitation density, implying a larger hot-carrier density due to efficient intraband scattering. These observations confirm the results of Ref.\ \cite{Tielrooij2013}, which showed linear scaling with photon energy (at very low fluence). \textcolor{blue}{\textbf{\ref{Fig4}B}} shows the measured and the theoretically calculated THz photoconductivity as a function of photon energy, at fixed fluence. The experimental and theoretical results are in good agreement (except for a constant scaling factor) and show a larger negative photoconductivity for larger photon energy. This effect is due to the higher electron temperature at larger photon energy. Due to saturation effects, the scaling with photon energy is not perfectly linear. This saturation likely comes from the reduced heating efficiency for increasing fluence and increasing photon energy, as has been observed in Ref.\ \cite{Jensen2014}. We will address this in more detail in \textcolor{blue}{\textbf{\ref{Fig8}}}.
\

\textcolor{blue}{\textbf{\ref{Fig4}C}} shows the calculated hot-carrier multiplication factor, defined as HCM = $(n_{\rm HC} - n_{\rm HC,0}) / n_{\rm exc}$, as a function of photon energy, obtained from our theoretical approach. For the investigated conditions, calculations show that the HCM factor is larger than 1 and increases with photon energy. For a photon energy of 2.5 eV and a fluence corresponding to a photoexcited carrier density of 0.2$\times$10$^{12}$/cm${^2}$, we calculate that 7 hot carriers are created on average per absorbed photon. This is fully compatible with the very recent experimental demonstration of 5 generated and subsequently collected hot carriers per absorbed photon in a photo-Nernst device \cite{Wu2016}. The plots of the electron distribution in \textcolor{blue}{\textbf{\ref{Fig5}}} illustrate the microscopic picture explaining hot-CM: excitation with a larger photon energy leads to a stronger broadening of the Fermi-Dirac distribution.
\

The panels in \textcolor{blue}{\textbf{\ref{Fig6}}} show the same quantities as in the corresponding panels of Fig.\ 4, but closer to the charge neutrality point (see also \textcolor{blue}{\textbf{\ref{Fig10}}}). Due to electron-hole puddles, the effective Fermi energy will not reach a value below 0.05--0.1 eV. In \textcolor{blue}{\textbf{\ref{Fig6}A}} and \textcolor{blue}{\textbf{\ref{Fig6}B}} we show the THz photoconductivity, measured for 3 photon energies $E_{\rm ph} = $ 0.5 eV, 1.0 eV, and 2.5 eV, as we vary the fluence of the pump pulse, compared to the results of the theoretical calculation. In the calculation we used $E_{\rm F}$ = 0.05 eV. The THz photoconductivity is now positive, with a larger signal for a larger photon energy. This suggests that interband heating is also an efficient process. We again observe saturation effects that are likely related to the decreased heating efficiency at higher fluence and photon energy (see Ref.\ \cite{Jensen2014} and \textcolor{blue}{\textbf{\ref{Fig8}}}). An additional effect giving rise to saturation could be a reduction of conductivity (leading to reduced positive photoconductivity) due to populating optical phonons at increasing fluence and photon energy \cite{Jnawali2013, Mihnev2016}.
\ 

In analogy to the case of intraband heating and hot-CM, in \textcolor{blue}{\textbf{\ref{Fig6}C}} we show the calculated carrier multiplication factor CM = $(n_{\rm CB} - n_{\rm CB,0}) / n_{\rm exc}$ as a function of the photon energy. Under the investigated conditions, we find that the CM factor is larger than 1 and increases with photon energy. For a photon energy of 2.5 eV and a fluence corresponding to a photoexcited carrier density of 0.2$\times$10$^{12}$/cm${^2}$, we calculate that $>$3 secondary carriers are created on average per absorbed photon. To put this in perspective, 2 electron-hole pairs are created per absorbed photon for bulk PbS excited with 3 eV photons \cite{Pijpers2009}. The plots of the carrier distribution in \textcolor{blue}{\textbf{\ref{Fig7}}} show stronger broadening of the Fermi-Dirac distribution for higher photon energy, corresponding to more electronic transitions from valence to conduction band. Thus, although the occurrence of CM arises from specific microscopic interactions, such as collinear scattering events \cite{Winzer2010, Winzer2012, Brida2013}, the thermodynamic picture of \textcolor{blue}{\textbf{\ref{Fig7}}} indicates that it can be regarded as interband heating with high efficiency. 
\

\subsubsection*{Heating efficiency}

Ultimately, both interband and intraband heating are mechanisms that slow down the dissipation of the electron energy density by phonons, and in this sense they have the same potential to increase the efficiency of a light-harvesting or photodetection device. In order to assess the ultimate potential of graphene-based photodetection, we characterize the efficiency of carrier heating, i.e.\ the energy transfer from photons to electrons, by calculating the carrier temperature at time $t = $ 0.1 ps, and the ratio between the electron energy density at time $t = $ 0.1 ps and the photon energy density of the pump pulse. We chose this delay time, because it corresponds to the quasi-equilibrium state, where the initial ultrafast energy relaxation, including interband and intraband heating, has already occurred, whereas equilibration with the phonon system is far from complete, i.e.\ the hot carrier distribution has been formed and has not cooled yet. \textcolor{blue}{\textbf{\ref{Fig8}}} shows these results for a photoexcited carrier density of 0.5 and 0.17$\times$10$^{12}$/cm$^2$ and 800 nm light. The calculations show that the quasi-equilibrium electron temperature is somewhat higher for lower carrier density (see \textcolor{blue}{\textbf{\ref{Fig8}A}}), mainly due the smaller electronic heat capacity. We also see that the higher fluence corresponds to a lower heating efficiency (see \textcolor{blue}{\textbf{\ref{Fig8}B}}), in agreement with the saturation effects in our measured photoconductivity vs. fluence (see \textcolor{blue}{\textbf{\ref{Fig4}}} and \textcolor{blue}{\textbf{\ref{Fig6}}}) and the results of Ref.\ \cite{Jensen2014}. The reason for the reduced heating efficiency is that at a higher fluence the energy transfer to phonons occurs faster (see \textcolor{blue}{\textbf{\ref{Fig8}C-D}}). Importantly, this thermodynamic picture shows that the energy transfer from photons to electrons is exceptionally efficient and depends only weakly on the equilibrium Fermi energy. This once more asserts the great intrinsic potential of graphene for photodetection applications, provided that hot carriers can be extracted before their cooling occurs (typically on a picosecond timescale at room temperature).
\

\section*{Discussion}

We have observed a correlation between, on the one hand, the transition from \textit{positive} to \textit{negative} THz photoconductivity, and on the other hand, the transition from \textit{interband} to \textit{intraband} heating, respectively. In the following, we discuss the physical origins behind positive and negative THz photoconductivity. There are two reasons to expect an increase of conductivity following photoexcitation (i.e.\ positive THz photoconductivity). First, the pump laser increases the electron density in the conduction band and generates a finite hole density in the valence band. The free carrier density in the system is larger and thus the conductivity increases.  Second, the generation of hot electrons entails an increase of the electron temperature. Picturing graphene as a zero-gap semiconductor, it is natural to expect that a higher temperature $T \simeq E_{\rm F} / k_{\rm B}$ corresponds to a larger conductivity as well. The reduction of conductivity (i.e.\ negative THz photoconductivity) is more subtle, as its origin resides in the reduction of electronic screening of the long-range Coulomb interaction between graphene's carriers and charged impurities in the substrate. More precisely, the polarizability function of graphene's carriers, evaluated at the Fermi wave vector where most scattering events take place, decreases with temperature and thus reduces the effectiveness of electronic screening \cite{Hwang2009}. When screening is less effective, impurities in the substrate interact more strongly with graphene's carriers and the conductivity decreases due to enhanced scattering.
\

The experimentally observed correlation between the transition from interband to intraband heating and from positive to negative THz photoconductivity can thus be explained as follows: In the vicinity of the charge neutrality point, where interband heating takes place, a larger free carrier density is available for conduction and this preempts the effect of the reduced screening at higher electron temperature. The THz photoconductivity is positive. At higher carrier density, where  intraband heating takes place, the system evolves to a regime where screening of the impurities in the substrate is less effective and the THz conductivity is decreased with respect to its equilibrium value. The THz photoconductivity is negative. Thus electronic screening plays a crucial role -- first, during the ultrafast dynamics that lead to a thermalized electron distribution \cite{Brida2013,Tomadin2013} and, second, for the resulting conductivity that corresponds to this thermalized distribution. Both CM and interband recombination due to carrier-carrier scattering vanish when a non-regularized dynamical screening model is used \cite{Brida2013,Tomadin2013}. Our calculations show that such a screening model yields positive THz photoconductivity at all equilibrium carrier densities, contrary to the experimental results, because it overestimates the free carrier density in the conduction band at large equilibrium carrier density.
\

To put our results in perspective, we point out that for semiconductors, the concept of CM is highly attractive, as it allows to overcome the Shockley-Queisser limit of conventional solar cells \cite{Shockley1961}, as well as photodetection with gain, where multiple electrons are detected per absorbed photon. Typical light harvesting and photodetection devices using semiconductors rely on the photovoltaic effect, where photoexcited electron-hole pairs are separated by the intrinsic electric field present at the junction between \textit{p}-doped and \textit{n}-doped regions. However, in graphene the photoresponse at a \textit{pn}-junction (e.g.\ created by local gates) is typically dominated by the photo-thermoelectric (PTE) effect \cite{Song2011, Gabor2011}, at least under zero bias voltage. Here the different Seebeck coefficients in the \textit{p} and \textit{n} regions give rise to an electron-heat driven photoresponse. It turns out that in this case efficient heating (and hot-CM) are highly beneficial for creating a large PTE response and indeed multiple hot carrier harvesting per absorbed photon has been demonstrated very recently in a high-responsivity photodetector \cite{Wu2016}. Furthermore, in a recent study, an increased photoresponse at a graphene \textit{pn}-junction photodetector (with a 3 dB bandwidth of 65 GHz) was observed upon increasing the bias voltage, which only occurred at low Fermi energy \cite{Schuler2016}. This observation was interpreted as arising from a photoconductive contribution to the photoresponse, coming from additional carriers in the conduction band, thus increasing the device conductivity. Our results explain why this effect only occurs at low Fermi energy: only in this regime does photoexcitation actually lead to additional carriers in the conduction band, i.e.\ $n_{\rm CB}>n_{\rm CB,0}$ (see Fig.\ 2C). In contrast, at higher Fermi energies $n_{\rm CB}\approx n_{\rm CB,0}$, so that no photoconductive effect is expected. The insights from our work lead to direct input for optimizing graphene-based photodetector devices. 
\

\section*{Methods}

\subsubsection*{Experimental design}

\begin{figure}[h!]
\centering
\includegraphics[scale=0.9]{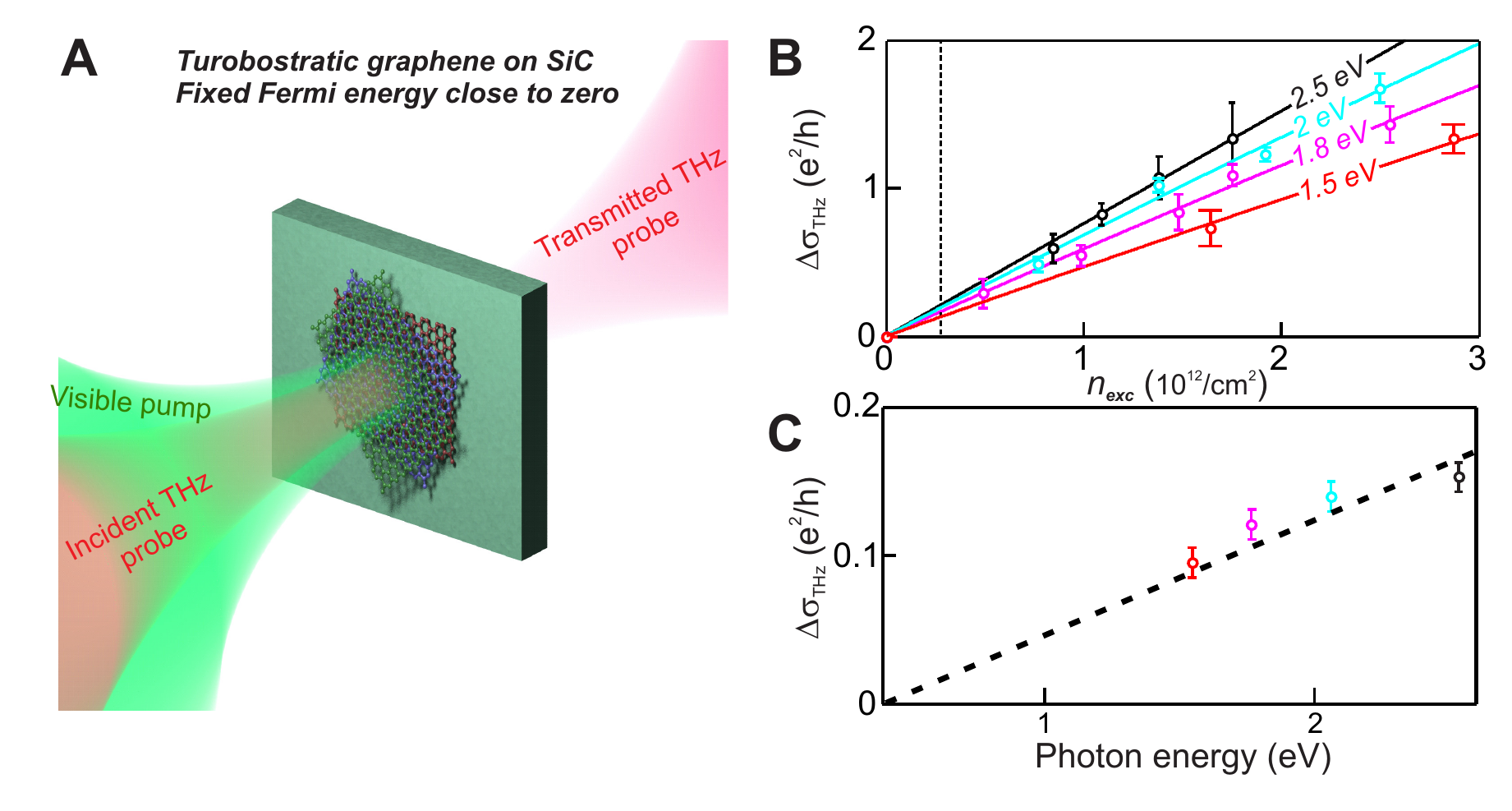}
\caption{
\textbf{A)} Illustration of the optical pump -- THz probe measurement technique applied to turbostratic graphene supported by SiC substrate (see Methods). This device consists of multiple graphene monolayers with random relative orientation and very low doping. Through these measurements, we further examine graphene close to the charge neutrality point, as in \textcolor{blue}{\textbf{\ref{Fig4}}} for the gate-tunable sample.
\textbf{B)} The THz photoconductivity $\Delta \sigma_{\rm THz}$ as a function of the pump pulse fluence, parametrized by the photoexcited carrier density $n_{\rm exc}$, for several values of the pump photon energy $E_{\rm ph} = $ 1.5, 1.8, 2.0, and 2.5 eV (empty circles). The turbostratic graphene samples are characterized by a very low doping, so this panel should be compared to \textcolor{blue}{\textbf{\ref{Fig4}A}}. The solid lines are linear fits to the data. \textbf{C)} The THz photoconductivity at $n_{\rm exc} = 0.2 \times 10^{12}$/cm$^2$ [vertical dashed line in \textbf{B)}], as a function of the pump photon energy. The dashed line is a linear fit to the data. Consistently with the THz photoconductivity measured on gated samples, we observe larger THz photoconductivity for larger photon energy, as discussed in \textcolor{blue}{\textbf{\ref{Fig4}}}. This trend indicates that efficient interband heating takes place in the, intrinsically undoped, turbostratic graphene samples. } \vspace{12pt}
\label{Fig10}
\end{figure}

For the gate-tunable device, we used CVD graphene (commercially grown by Graphene Supermarket), which was transferred onto a quartz substrate. Quartz is used as opposed to the more common silicon substrate for its negligible photoconductivity and high THz transmission. The ionic gate consists of LiClO$_4$ mixed with PEO (Poly Ethyl Oxide) in a ratio 8:1 by mass. This mixture is dissolved in methanol to apply it to the graphene. The source and drain contacts were made from 100 nm Au and 5 nm Cr. Turbostratic graphene \cite{Pimenta2007, Geim2009} (see \textcolor{blue}{\textbf{\ref{Fig10}}}) was obtained on the C-face of SiC through thermal decomposition in argon atmosphere. The graphene thickness was estimated to be about 8 layers via Raman spectroscopy  \cite{Candini2015, Convertino2016}. The optical pump -- THz probe setups that were used to measure THz photoconductivity are very similar to the one described in Ref.\ \cite{Ulbricht2011}, the main difference being that the pump path contains an optical parametric amplifier and mixing stages (Light Conversion - Topas) to convert the incident 800 nm light to different wavelengths. We measured the pump-induced change in THz transmission at the peak of the pump-probe time trace (see \textcolor{blue}{\textbf{\ref{Fig9}A}}), with the 800 nm sampling beam (which measures the THz pulse through electro-optic sampling) positioned at the peak of the quasi-single cycle THz pulse. In this configuration, the pump-induced change in THz absorption is measured, corresponding to the pump-induced change in the real part of the conductivity, as extracted using the thin film approximation (see e.g. Ref.\ \cite{Ulbricht2011, Tielrooij2013}).  
\

\subsubsection*{Calculation of the THz conductivity}

The expression for the intraband conductivity $\sigma(t)$ in graphene at angular frequency $\omega$ reads \cite{Hwang2009}
\begin{equation}\label{eq:conductivity}
\sigma_{\rm THz}(t) = -\frac{e^{2} v_{\rm F}}{2} \sum_{\lambda} \int_{0}^{\infty} d\varepsilon \nu(\varepsilon) \frac{\tau(\varepsilon;t)}{1 - i \omega \tau(\varepsilon;t)} \frac{\partial f_{\lambda}(\varepsilon; t)}{\partial \varepsilon}~,
\end{equation}
where $-e$ is the electron charge, $v_{\rm F}$ the Fermi velocity in graphene, $\nu(\varepsilon) = 2 |\varepsilon| / [\pi(\hbar v_{\rm F})^{2}]$ the density of states, $\lambda=0,1$ labels electron states in the conduction band or hole states in the valence band, and $f_{\lambda}(\varepsilon; t)$ is the corresponding quasi-equilibrium distribution function at energy $\varepsilon > 0$ (measured from the Dirac point) and time $t$. The quantity $f_{\lambda}(\varepsilon; 0)$ represents the equilibrium distribution before photoexcitation. The scattering mechanism responsible for the finite conductivity determines the transport scattering time $\tau(\varepsilon;t)$, which depends on the distribution function. Eq.~(\ref{eq:conductivity}) is justified as long as $\hbar \omega \ll 2 E_{\rm F}$, such that the probe pulse does not produce inter-band electron transitions. The expression for the transport scattering time due to scattering with long-range Coulomb impurities reads \cite{Hwang2007}
\begin{equation}\label{eq:scatteringtime}
\frac{\hbar}{\tau(\varepsilon;t)} = \pi n_{\rm i} \sum_{{\bm k}'} |W_{\lambda}(q;t)|^{2} (1 - \cos^{2}{\theta_{{\bm k}'}}) \delta(\varepsilon_{{\bm k}'} - \varepsilon),
\end{equation}
where $n_{\rm i}$ is the impurity density, $\theta_{{\bm k}'}$ is the polar angle of the wave vector ${\bm k}'$, $q$ is the modulus of the difference between ${\bm k}'$ and ${\bm k} = \hat{\bm x} \varepsilon / (\hbar v_{\rm F})$, and $\varepsilon_{\bm k} = \hbar v_{\rm F} k$ is the dispersion of electrons and holes. We denote by $W_{\lambda}(q;t)$ the screened Coulomb potential, that reads
\begin{equation}\label{eq:screenedcoulomb}
W_{\lambda}(q;t) = (-1)^{\lambda} \frac{v_{q}}{\epsilon(q,0;t)}, \quad v_{q} = \frac{2 \pi e^{2}}{\bar{\epsilon} q} ~,
\end{equation}
where $\bar{\epsilon}$ is the average dielectric constant around the graphene plane, $\epsilon(q,0;t)$ is the static electronic dielectric function in graphene, and the Coulomb impurities have been taken to sit at the graphene plane for simplicity. In the well-known random phase approximation (RPA), the dielectric function is related to the polarizability function $\chi^{(0)}(q,\omega;t)$ of the non-interacting electron system (i.e.~the Lindhard function) by $\epsilon(q,\omega;t) = 1 - v_{q} \chi^{(0)}(q,\omega;t)$. The expression for the Lindhard function of the thermalized photoexcited electron system in graphene has been obtained in Ref.\ \cite{Tomadin2013}. The Lindhard function depends on the electron temperature $T(t)$ and the chemical potentials $\mu_{\rm e}(t)$, $\mu_{\rm h}(t)$ for electrons in the conduction band and holes in the valence band at time $t$.  In our approach, we obtain these quantities by integrating in time the semiclassical Boltzmann equation for the coupled electron-optical phonon system, as detailed in Refs.\ \cite{Brida2013,Tomadin2013}. Then, at each time $t$ of interest, we calculate the transport scattering time on an energy mesh with Eq.~(\ref{eq:scatteringtime}) and, finally, the conductivity with Eq.~(\ref{eq:conductivity}). We remark that this approach takes {\it screening} into account, within the RPA approximation, at each instant in time, without making assumptions on the energy dependence of the transport scattering time.
\

\subsection*{Acknowledgements}
We thank Zolt\'{a}n Mics, Justin Song, Nils Richter and Paolo Fantuzzi for fruitful discussions and Fabien Vialla for graphical support. This work was supported by the European Union's Horizon 2020 research and innovation programme under grant agreement No. 696656 “Graphene Flagship”, the EU grants FP7-ICT-2013-613024-GRASP and Moquas FET-ICT-2013-10 \#610449, Fondazione Istituto Italiano di Tecnologia, the Spanish Ministry of Economy and Competitiveness through the “Severo Ochoa” Programme for Centres of Excellence in R\&D (SEV-2015-0522), Fundacio Cellex Barcelona, the  Mineco grants Ramon y Cajal (RYC-2012-12281), Plan Nacional (FIS2013-47161-P), the Government of Catalonia trough the SGR grant (2014-SGR-1535), the ERC StG “CarbonLight” (307806), the German Research Foundation (DFG Priority Program Graphene SPP 1459, KL1811), as well as the State Research Centre for Innovative and Emerging Materials (CINEMA) and the Graduate School of Excellence Materials Science in Mainz (MAINZ) GSC 266. K.J.T. acknowledges support through the Mineco Young Investigator Grant (FIS2014-59639-JIN). E.H. acknowledges support from EPSRC fellowship EP/K041215/1. 
\\

\onecolumngrid

\clearpage

\section*{Additional figures}

\end{document}